\documentclass[aps,prl,reprint,showpacs,showkeys,superscriptaddress]{revtex4-1}
\usepackage{multirow}
\usepackage{graphicx}
\usepackage{bm}
\usepackage{dcolumn}
\usepackage[colorlinks=true,linkcolor=blue,urlcolor=blue,citecolor=blue]{hyperref}
\usepackage{natbib}
\usepackage{amsmath}
\usepackage{times}

\begin{document}

\preprint{2}

\title{Effects of quantum impurity spins on the magnetic properties of zigzag and linear spin chains}
\author{Koushik Karmakar}\affiliation{Indian Institute of Science Education and Research, Pune, Maharashtra-411008, India}
\author{Markos Skoulatos}\affiliation{Laboratory for Neutron Scattering and Imaging, Paul Scherrer Institute, 5232 Villigen, Switzerland}\affiliation{Physik Department, Technische Universit\"at M\"unchen, D-85748 Garching, Germany}
\author{Giacomo Prando}\affiliation{Center for Transport and Devices of Emergent Materials, TU Dresden, D-01062 Dresden, Germany}\affiliation{Leibniz-Institut f\"ur Festk\"orper- und Werkstoffforschung (IFW) Dresden, D-01171 Dresden, Germany}
\author{Bertran Roessli}\affiliation{Laboratory for Neutron Scattering and Imaging, Paul Scherrer Institute, 5232 Villigen, Switzerland}
\author{Uwe Stuhr}\affiliation{Laboratory for Neutron Scattering and Imaging, Paul Scherrer Institute, 5232 Villigen, Switzerland}
\author{Franziska Hammerath}
\affiliation{Leibniz-Institut f\"ur Festk\"orper- und Werkstoffforschung (IFW) Dresden, D-01171 Dresden, Germany}
\author{Christian R\"uegg}\affiliation{Laboratory for Neutron Scattering and Imaging, Paul Scherrer Institute, 5232 Villigen, Switzerland}\affiliation{Department of Quantum Matter Physics, University of Geneva, 1211 Geneva, Switzerland}
\author{Surjeet Singh}\email[email:]{surjeet.singh@iiserpune.ac.in}\affiliation{Indian Institute of Science Education and Research, Pune, Maharashtra-411008, India}
\date{\today}

\begin{abstract}

We investigated the magnetic ground state and low-energy excitations of the spin chains compounds SrCuO$_{2}$ (zigzag chains) and Sr$_{2}$CuO$_{3}$ (linear chains) in the presence of quantum impurities induced by lightly doping ($\leq 1 \%$) with Zn$^{2+}$ ($S = 0$), Co$^{2+}$ ($S =1/2$) and Ni$^{2+}$ ($S = 1$) impurities at the Cu$^{2+}$ site. We show that the ground states and the nature of low-lying excitations (i.e., gapped or gapless) depend on the spin state and symmetry of the defects. For Ni doped chains a spin gap is observed but for Zn and Co doping the excitations remain gapless. Co-doped chains exhibit magnetic order with critical temperatures significantly enhanced compared to those of the pristine compounds. In the specific case of 1 \% Co impurities, the linear chains exhibit long-range order below 11 K, while the zigzag chain is characterized by a quasi-long range ordered phase below 6 K with correlation lengths of about 12\textit{a} and 40\textit{c} units along the crystal axes \textit{a} and \textit{c}, respectively. The different magnetic behaviours of these two compounds with comparable intra- and interchain couplings underpin the role of spin frustration in the zigzag chains.

\end{abstract}


\maketitle

\textit{Introduction:} Quantum spin chains exhibit several novel properties related to the nature of their low-lying spin excitations. These are known to be \textit{gapless} for chains comprising half-integral spins and gapped for the integral spins \cite{Haldane1983}. Due to spatial confinement of spinons, spin chains are expected to be highly sensitive to the presence of defects or disorder. There has been considerable theoretical work on the random Heisenberg antiferromagnetic (HAF) spin chains in the strong-disorder limit, pioneered by Ma, Dasgupta and Hu using the renormalization group (RG) methods \cite{Ma1979, Refael2013}. The strong-disorder RG theory predicts a stable fixed point, called the random-singlet (RS) phase, to which the system flows to irrespective of the details of the disorder. Recent experimental works, however, indicate a similar universal behaviour in the \textit{weak-disorder} limit. More specifically, low level of Ca doping ($\sim 10$ \%) at the Sr site outside the chain structure in the spin chain compounds SrCuO$_{2}$ and Sr$_{2}$CuO$_{3}$ results in opening of a spin \textit{pseudogap} in their low-lying excitation spectra, arising from the bond disorder within the chains due the ionic size mismatch of Sr$^{2+}$ and Ca$^{2+}$ ions \cite{Hammerath2011, Hammerath2014}. Similarly, merely 1 \% Ni impurities at the Cu-site in SrCuO$_{2}$ opens up a sizeable gap ($\sim 8$ meV), not related to any structural transition. These observations led Simutis \textit{et al.} to suggest that the spin pseudogap is a generic feature of quantum spin chains with dilute defects \cite{Simutis2013}. Here, we investigate this point further by considering low ($\leq 1 \%$) concentrations of defects characterized by different spin quantum numbers ($S$). The role of dilute defects is also relevant to such diverse areas as quantum computing \cite{Bose2007} and heat management \cite{Otter2012} where quantum spin chains are expected to play a crucial role.

The main focus of the present letter is to address these issues in the prototype spin chain systems SrCuO$_{2}$ (\textit{zigzag} chains) and Sr$_{2}$CuO$_{3}$ (\textit{linear} chains), which have an orthorohmbic crystal structure with chains along the $c-$ and $b-$ axis, respectively. The disorder is induced by doping the chains at the Cu$^{2+}$ site with Zn$^{2+}$ ($S = 0$), Co$^{2+}$ ($S = 1/2$) and Ni$^{2+}$ ($S = 1$) impurities. By means of inelastic neutron scattering (INS), we show that the ground state of Ni-doped Sr$_{2}$CuO$_{3}$ is gapped, in line with the previous INS study on similar Ni-doped SrCuO$_{2}$ \cite{Simutis2013}. On the other hand, the ground state of Co- and Zn- doped crystals is found to be gapless. We therefore show that in the weak-disorder limit the details of disorder cannot be overlooked, as opposed to the strong-disorder case where a \textit{random singlet} phase is always favoured. Our measurements of elastic neutron scattering (ENS) and muon-spin spectroscopy ($\mu^{+}$SR) further evidence that the Co-doping promotes a quasi-long-range ordered (LRO) magnetic phase in SrCuO$_{2}$ whose critical temperature increases with increasing Co content. The role of \textit{zigzag} geometry of the spin chains in preventing the Co-doped SrCuO$_{2}$ from attaining a \textit{true} long-range ordered state is discussed. 

\textit{Experimental details}: All the experiments reported here were performed on high quality single crystals with mosaic spread of $< 0.5^{\circ}$, grown using the travelling-solvent-floating-zone method. Scanning electron microscopy and optical microscopy under polarized light were performed to confirm homogeneous doping in the crystals. The actual dopant concentration in the Co and Zn doped crystals was found to be almost half the nominal composition. In the Ni case, the measured and the nominal concentrations were almost the same. For details of the crystal growth see refs. \citenum{Karmakar2014, Karmakar2015b}. The dopant concentrations in the crystals were independently confirmed using susceptibility analysis \cite{Karmakar2015a}. Here, we have indicated the nominal compositions in each case. The neutron scattering experiments were performed at SINQ (Paul Scherrer Institute, Switzerland) on oriented single crystals weighing about $2$ g. The INS experiments were performed at $T = 1.5$ K using the thermal triple-axis spectrometer EIGER with final neutron energy $E_{f} = 14.7$ meV filtered using a graphite filter, while the ENS experiments were carried out on the cold triple-axis spectrometer RITA-II, with final neutron energy $E_{f} = 5$ meV, using a Be filter to remove the $\lambda/2$ contamination. Measurements of muon-spin spectroscopy ($\mu^{+}$SR) were performed at S$\mu$S (PSI) on the GPS spectrometer ($\pi$M3 beamline) in the temperature region 1.6 K $< T <$ 20 K, in conditions of zero-magnetic field (ZF$-\mu^{+}$SR) \cite{Blundell1999, Yaouanc2011}. The dc magnetization and ac magnetic susceptibility were measured using a Physical Property Measurement System (Quantum Design, USA).

\begin{figure}[t!]
\includegraphics[width = 8.5cm]{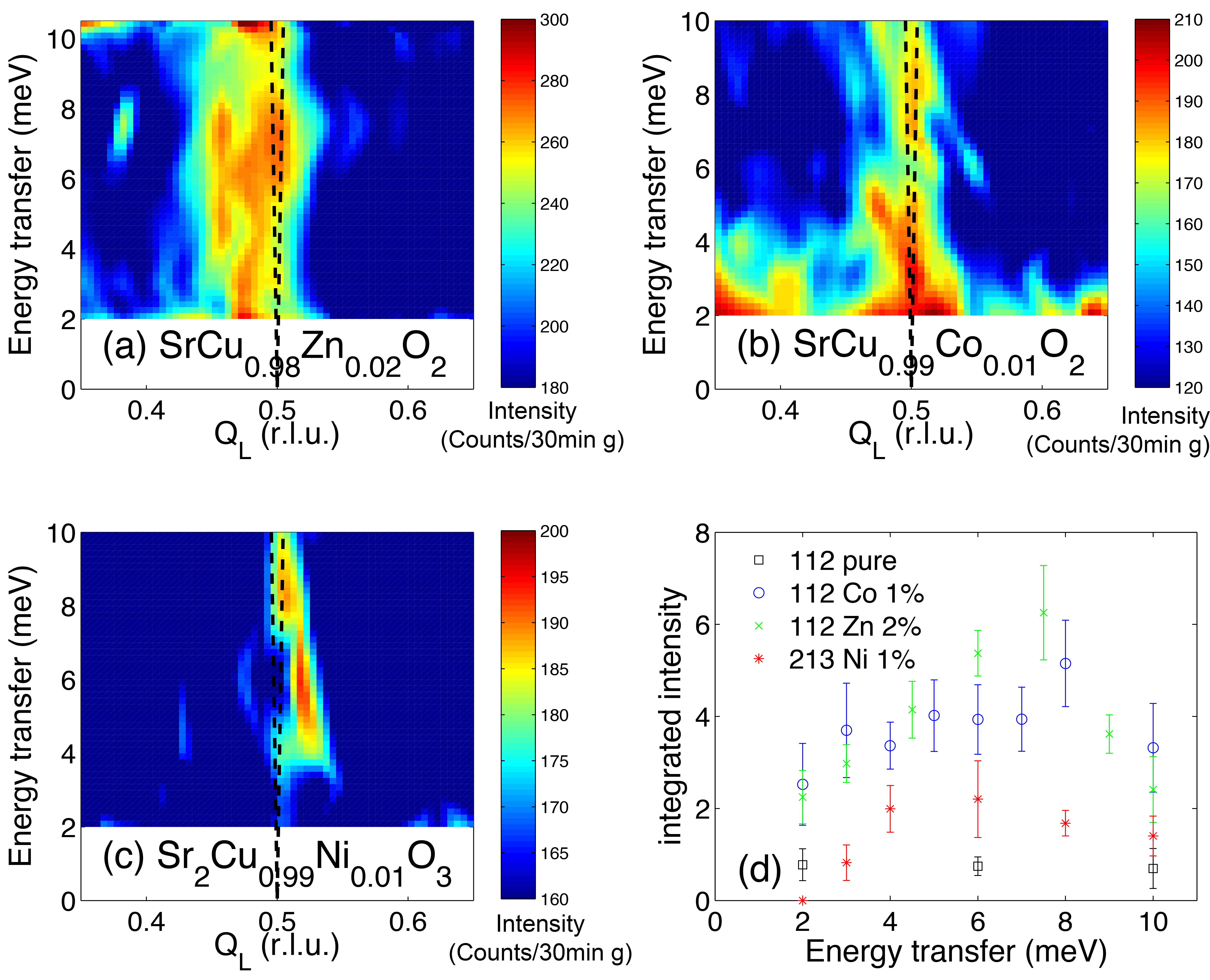}
\caption{Intensity colour maps of the low energy excitation spectra for various doped spin chains. The dashed lines correspond to the lower spinon boundary according to the `des Cloizeaux-Pearson' analytical result \cite{DesCloizeaux1962}. Panels (a)-(c) correspond to Zn-, Co- and Ni-doped cases respectively while panel (d) shows the corresponding integrated neutron intensities as a function of energy transfer. Undoped compound is measured for comparison.}
\label{INS}
\end{figure}

\textit{Results}: INS experiments are summarized in Fig. \ref{INS}, where panels (a) and (b) correspond to Zn- and Co- doped SrCuO$_{2}$, respectively, and panel (c) to Ni doped Sr$_{2}$CuO$_{3}$. The vertical dashed lines originating at $Q_{L} = 1/2$ indicate the des Cloiseaux-Pearson characteristic dispersion of spinon excitations \cite{DesCloizeaux1962}, for intra-chain coupling, $J = 230$ meV \cite{Rosner1997}. Panel (d) shows the normalized \textit{Q}-integrated intensities as a function of energy transfer up to 10 meV for all the considered samples. An undoped SrCuO$_{2}$ single crystal has been also measured as a reference. Surprisingly, all the doped samples have average spectral weights that are almost five-fold enhanced in this energy range as compared to the pure sample. However, what is interesting here is that the intensities of the Co- and Zn-doped crystals show little variation as a function of the energy transfer, indicating that these samples are gapless down to the instrumental energy resolution of 2 meV. Contrary to this, the Ni-doped samples [Fig. \ref{INS}(c)] evidences a gapped behaviour with considerably reduced intensity below $\sim 4$ meV. 

\begin{figure}[t!]
\includegraphics[width = 8.5 cm]{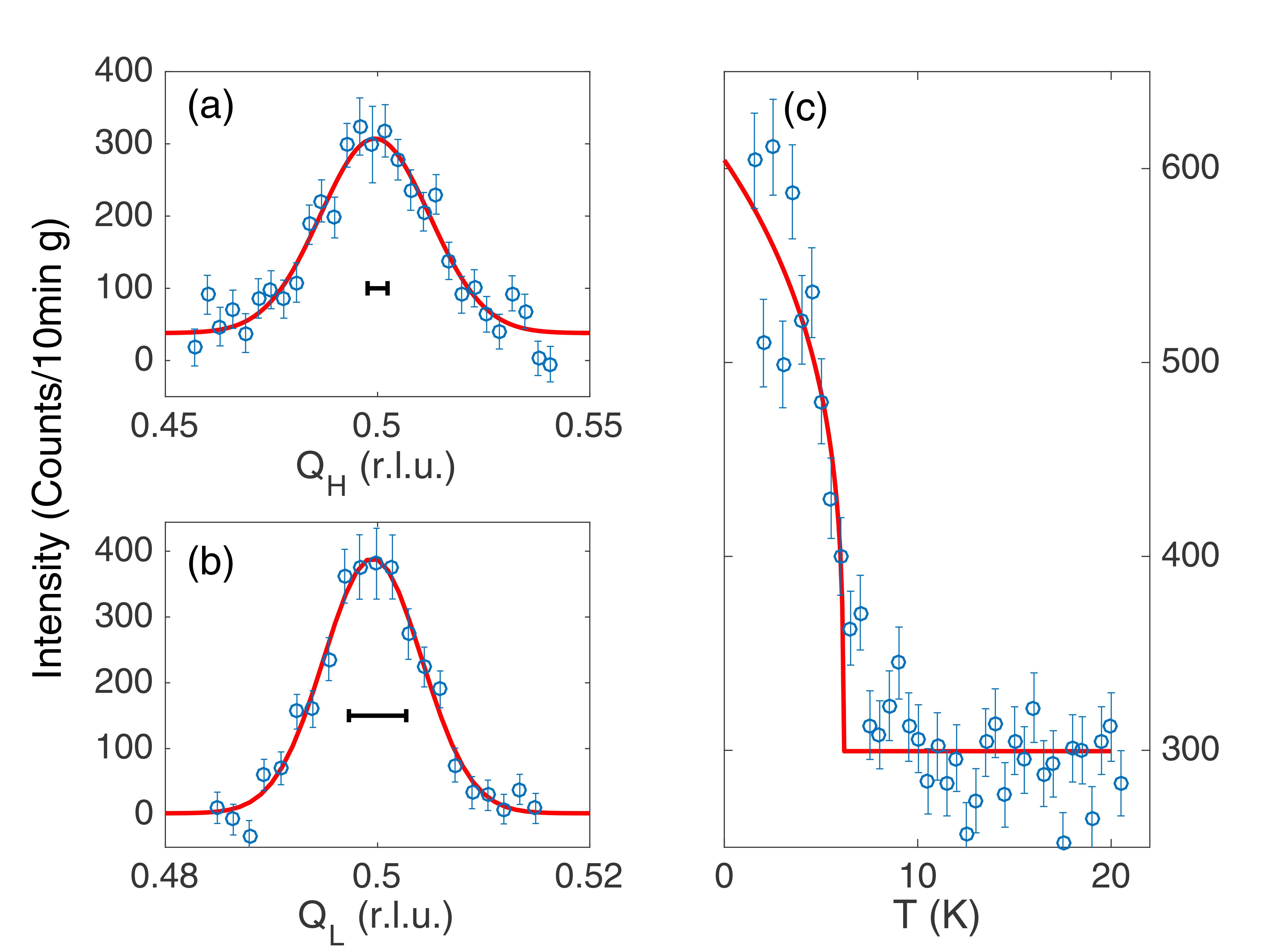}
\caption{Magnetic scattering of SrCu$_{0.99}$Co$_{0.01}$O$_{2}$ at the (1/2 0 1/2) Bragg peak. (a) and (b) depict the $Q_{H}$ and $Q_{L}$ dependence at 1.5 K. Solid lines are Gaussian fits, while the horizontal bars refer to the instrumental resolution. (c) shows the order parameter as a function of T, with a power-law fit (solid line) yielding $T_{N} = 6.1(3)$ K.}
\label{ENS}
\end{figure}

ENS data for Sr$_{0.99}$Co$_{0.01}$CuO$_{2}$ are presented in Fig. \ref{ENS}, where the magnetic scattering around $(1/2, 0, 1/2)$ at $T = 1.5$ K is displayed. Measurements were also performed at $T = 20$ K, allowing us to subtract a weak, higher-order wavelength contamination. Panels (a) and (b) show the background-subtracted pure magnetic scattering along the $Q_{H}$ and $Q_{L}$ reciprocal lattice vectors, respectively. These peaks are broadened in both directions uncharacteristic of a purely LRO magnetic phase, which is characterized by instrumental-resolution-limited peak widths. The corresponding correlation lengths obtained are $\xi_{a} = 44$ \AA\, ($\sim 12\, a$) and $\xi_{c} = 155$ \AA\, ($\sim 40\, c$). Finally, panel (c) shows the temperature dependence of the magnetic Bragg peak, with a power-law fit yielding $T_{N} = 6.1(3)$ K. Accordingly, our ENS results suggest that SrCu$_{0.99}$Co$_{0.01}$O$_{2}$ crystals exhibit an anisotropic quasi-LRO magnetic phase below $T \approx 6$ K. It should be remarked that due to Co-doping the $T_{N}$ is sizeably enhanced from a value of $< 2$ K in the pristine compound \cite{Matsuda1997}.

\begin{figure}[t!]
\vspace*{-0.4cm}
\includegraphics[width = 8.2cm]{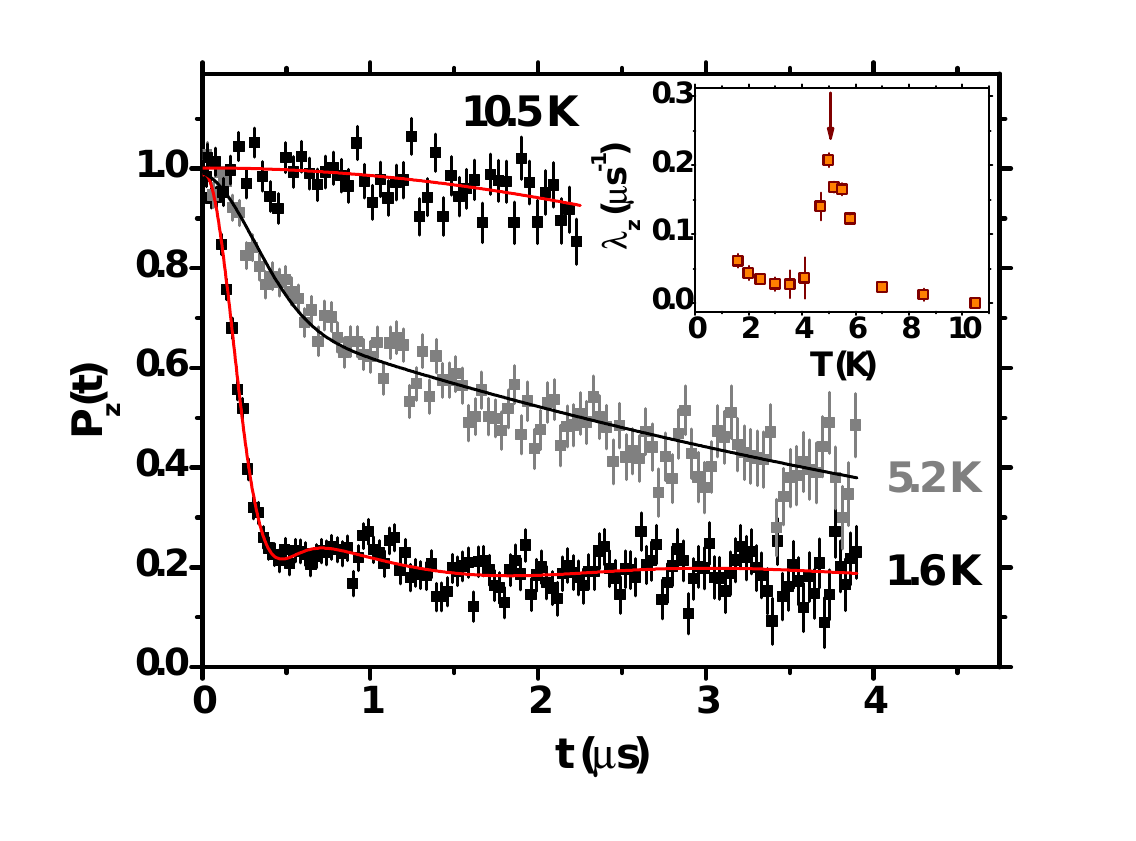}
\vspace*{-0.4cm}
\caption{(De)polarization ($P_{z}$) as a function of time (t) for representative temperatures ($T = 1.6, 5.2 and 10.5$ K), shown in ZF conditions. Continuous lines are best-fits to the experimental data according to Eq. \ref{EqMuons}. Inset: $T$ dependence of the longitudinal relaxation rate. The arrow indicates the estimated $T_{N}$ value.}
\label{Figure} 
\end{figure}

To further characterize the magnetic phase we use ZF-$\mu^{+}$SR \cite{Blundell1999,Yaouanc2011}, where the spin (de)polarization function $P_{z}(t)$ of implanted $\mu^{+}$ is measured as a function of time ($t$). In our measurements \cite{Prando2016}, the samples were aligned in order to have the $z$ direction parallel to the crystallographic $b$ axis, but with an arbitrary orientation of the $a$-$c$ plane. Experimental $P_{z}(t)$ curves for SrCu$_{0.99}$Co$_{0.01}$CuO$_{2}$ are reported at representative $T$ values in Fig. \ref{Figure}. Curves were fitted using the function: 
\begin{equation}\label{EqMuons}
P_{z}(t) = \sum_{i} a_{\text{T}_{i}} \cos\left(\gamma B_{i} t\right) e^{-\sigma_{\text{T}_{i}}^{2}t^{2}/2} + a_{\text{L}} e^{-\lambda_{\text{L}}t}
\end{equation}
usual for materials where a LRO magnetic phase is expected to develop \cite{Prando2013}. Here, $\lambda_{\text{L}}$ is the longitudinal spin-lattice relaxation rate. Its $T$ dependence is shown in the inset of Fig. \ref{Figure}, displaying a narrow critical peak around the magnetic transition. At the same time, the development of two ($i = 1,2$) highly damped, though visible, oscillations in the transversal fraction at low $T$ denotes a local homogeneous magnetic environment for the $\mu^{+}$ whose distribution width $\sim \sigma_{\text{T}}$ is still smaller than the average $B$ value. These features further corroborate that the magnetic ground state of the system as a quasi-LRO phase. Finally, in agreement with ENS data, orienting $z$ along the $a$-$c$ plane results in an anisotropic behaviour of $P_{z}(t)$, which will be discussed in detail elsewhere \cite{Prando2016}.

\begin{figure}[t!]
\includegraphics[width = 8.2 cm]{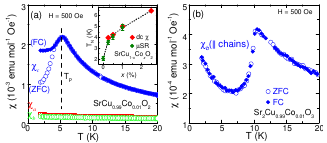}
\includegraphics[width = 8.2 cm]{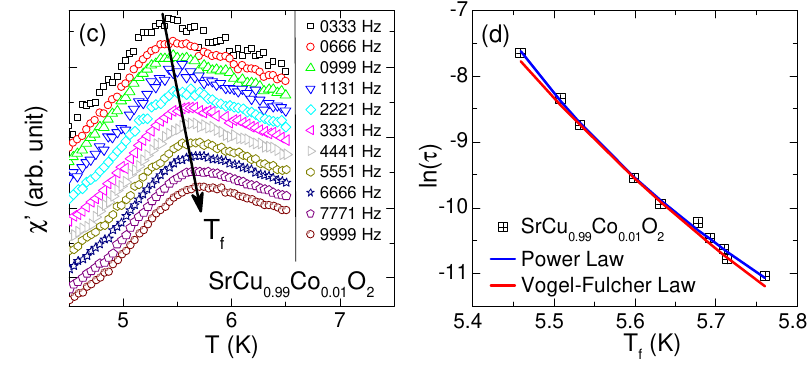}
\caption{Temperature dependent magnetic susceptibility of (a) SrCu$_{0.99}$Co$_{0.01}$O$_{2}$ and (b) Sr$_{2}$Cu$_{0.99}$Co$_{0.01}$O$_{3}$. Inset in panel (a) shows the variation of $T_{N}$ with Co concentration. (c)Temperature dependence of the real part of ac magnetic susceptibility ($\chi'$). (d) Relaxation time ($\tau$) with respect to the peak temperatures. The solid lines are fit to the data (see text for details). The curves in panel (c) are down shifted for clarity.}
\label{DC_CHI}
\end{figure}

Fig. \ref{DC_CHI}(a) shows the temperature dependence of the dc magnetic susceptibility ($\chi$) for Sr$_{0.99}$Co$_{0.01}$CuO$_{2}$. $\chi$ along the crystallographic \textit{a} and \textit{b} axes ($\chi_{a}$ and $\chi_{b}$, respectively) is comparable in magnitude with that of the pristine compound \cite{Karmakar2014}. On the other hand, $\chi_{c}$ (parallel to the chains) is significantly enhanced and shows a well-defined peak at $T = 5.4$ K. In the ordered state, a splitting of zero-field cooled (ZFC) and field-cooled (FC) branches was also observed. We notice that $T_{N}$ is strongly dependent on the Co concentration, as shown in the inset of Fig. \ref{DC_CHI}(a) where data from both dc magnetisation and $\mu$SR \cite{Prando2016} are shown. As comparison, $\chi$ for Sr$_{2}$Cu$_{0.99}$Co$_{0.01}$O$_{3}$ parallel to the chain is shown in Fig. \ref{DC_CHI}(b). The data exhibit a sharp peak near $T = 11$ K which marks the onset of a magnetic LRO phase (confirmed using specific heat measurements not shown here). It is interesting to note that no thermomagnetic irreversibility in the susceptibility of linear chains could be detected down to the lowest temperature of $2$ K in our experiments, which is a crucial difference compared to the zigzag chains under identical doping conditions. Below $T = 4.5$ K, we observe an upturn likely related to a second phase transition at even lower temperatures.

Fig. \ref{DC_CHI}(c) shows the temperature dependence of the real part ($\chi_{ac}^{\prime}$) of the ac magnetic susceptibility, measured by applying an alternating field $H_{ac} = 5$ Oe with frequency (\textit{f}) ranging from $100$ Hz to $10000$ Hz. $\chi_{ac}^{\prime}$ signals at frequencies smaller than $100$ Hz were extremely weak, and the imaginary part $\chi_{ac}^{\prime\prime}$ was even weaker with poor signal-to-noise ratio over the entire frequency range.$\chi_{ac}^{\prime}$ also exhibits a peak in correspondence to the peak observed in dc $\chi$. The peak position shifts towards higher temperatures with increasing \textit{f}, as is commonly observed in glassy systems \cite{Charilaou2011}. The corresponding ratio $\Delta T_{f}/T_{f}$ was found to be around 0.04 per $\omega$ decade ($\omega = 2 \pi f$). This value is higher than what is typically reported for atomic spin glasses but smaller than the corresponding values reported for superparamagnets \cite{Mydosh1993}. A more quantitative analysis can be performed by plotting $ln(\tau)$, where $\tau = (2 \pi f)^{-1}$ is the relaxation time, against the freezing temperature ($T_{f}$) for various ac excitation frequencies [Fig. \ref{DC_CHI}(d)]. The data points were fitted using the theory of dynamical scaling near a phase transition \cite{Mathieu2004}: $\tau = \tau_{0}\left(\frac{T_{f}}{T_{g}}-1\right)^{-z\nu}$ , where $\tau_{0}$ represents a microscopic characteristic time, $T_{g}$ is the spin glass temperature in the limit $f \rightarrow 0$ and $z\nu$ denotes the critical dynamical exponent. From the fitting, we get a value of $\tau_{0} \sim 10^{-10 \pm 1}$ s, $z\nu \sim 4 \pm1,$ and $T_{g} = 5.2 \pm 0.1$ K. The value of $T_{g}$ is in good agreement with the position of the peak in the dc magnetic susceptibility. For atomic spin-glasses, the values of $\tau_{0}$ and of $z\nu$ typically lie in the range $10^{-13 \pm 1}$ s and $8 \pm 1$, respectively \cite{Mathieu2004}. The values obtained here fall in the range that is more frequently reported for systems with inhomogeneous or short-range ordering \cite{Sniadecki2011, Hanasaki2007, Lu2010, Wang2009, De2007}. An analogous analysis can be performed using the Vogel-Fulcher law $\tau = \tau_{0} \exp\left(\frac{E_{a}/k_{B}}{T-T_{0}}\right)$, which is typically used for glassy systems comprising short range order, yielded a reasonably good fit by fixing the Vogel temperature $T_{0}(<T_{g})$ and varying $\tau_{0}$ and the energy barrier $(E_{a})$ as fitting parameters. For $T_{0} = 4.2$ K, the values obtained for $\tau_{0}$ and $E_{a}$ are respectively $10^{-10 \pm 1}$ s and $20 \pm 2$ K. The value of the former parameter is in good agreement with the dynamical scaling approach.

\textit{Discussion}: Antiferromagnetically coupled Ni$^{2+}$ ($S = 1$) defects in a $S = 1/2$ HAF  spin chain are argued to be screened in a manner analogous to the Kondo-like screening of a impurity spin in a non-magnetic metallic host \cite{Eggert1992}. Such an extended defect site essentially decouples the chain segments, thereby confining the spinons to finite-length chain sections of average length $\langle L\rangle$, which yields a pseudogap in the spinon excitation, whose size roughly scales as inverse of $\langle L \rangle$. Recent NMR results support this hypothesis by showing that the spin pseudogap scales with the impurity concentration \cite{Utz2015}. The appearance of a gap is also confirmed independently through the analysis of bulk susceptibility data \cite{Karmakar2015a}. Contrary to this, no detectable gap is found in the crystals doped with Zn or Co defects, down to $2$ meV. In the case of scalar defects (Zn$^{2+}$, $S = 0$), chain segments across the defect may be bridged by an appreciable next-nearest-neighbour interaction ($J' \approx 140$ K \cite{Rosner1997}). The spinons are, therefore, only weakly confined, which is not sufficient to produce a measurable spin gap. The Zn-doping should, however, lower the ordering temperature, as is indeed found to be the case \citep{Karmakar2015a}. This behaviour is analogous to that of Pd ($S = 0$) doped Sr$_{2}$CuO$_{3}$ crystals \cite{Kojima1997} and is in accordance with the quantum mechanical calculations of ref. \citenum{Eggert2002}. 

Co-defects also appear to favour a magnetically ordered ground state albeit with an enhanced $T_{N}$ and highly anisotropic magnetisation, which is presumably related to the Ising-like spin of Co$^{2+}$ ions \cite{Bera2014, Slonczewski1961}. In Sr$_{2}$CuO$_{3}$, for example, the $T_{N}$ is increased from $5.4$ K in the pristine sample to $11$ K in the sample doped with 1 \% Co. The behaviour of SrCuO$_{2}$ upon Co-doping is analogous, with the exception that the ordering temperature is slightly lower ($T_{N} \approx 6$ K) and the ordered phase is not truly LRO. This lack of LRO is related to the zigzag geometry of its chains. Though interchain interactions favour LRO of the spins (as in Sr$_{2}$CuO$_{3}$), frustrated interactions in the zigzag chain tends to suppress the LRO below $T = 2$ K\cite{Matsuda1997, Haldane1988}. If, however, spin disorder is also included in the zigzag chains, a glassy phase will be favoured \cite{Zaliznyak1998}. It is interesting that similar levels of Co$^{2+}$ defects in the linear and zigzag chains of comparable intra- and interchain exchange interactions exhibit different behaviours, which underpins the role played by the spin frustration in chains with a zigzag geometry.

\textit{Conclusion}: In summary, we investigated the low-energy spin excitations of the spin chain compounds SrCuO$_{2}$ (zigzag chains) and Sr$_{2}$CuO$_{3}$ (linear chains) in the presence of weak disorder, introduced by doping Zn ($S = 0$), Co ($S = 1/2$), and Ni ($S = 1$). We show that Ni doping in Sr$_{2}$CuO$_{3}$ results in a spin gap analogous to the pseudogap of Ni-doped SrCuO$_{2}$ reported previously. On the other hand, doping with either Zn or Co in these compounds does not produce a spin gap as their spin excitation spectra remain gapless down to 2 meV energy transfer. Our experiments, therefore, show that whether or not the ground state of a weakly doped spin $1/2$ HAF will be gapped depends on the quantum state and symmetry of the doped defect. For the zigzag chains in SrCuO$_{2}$, we show that doping with Co for Cu results in a nearly LRO spin state below $T \approx 6$ K. The magnetic correlation lengths were found to be about 40 and 12 unit cell dimensions along and perpendicular to the chains. $\mu$SR and bulk magnetic susceptibilities corroborate these observations. Our investigations suggest that the weakly doped spin $1/2$ HAF exhibit a rich variety of magnetic ground states which should invite further experimental and theoretical investigations.

\textit{Acknowledgement}: S. S. And C.R. acknowledge financial support under project no.: INT/SWISS/ISJRP/PEP/P-06/2012. M. S. acknowledges funding from the European Community's Seventh Framework Programme (FP7/2007-2013) under grant no. 290605 (PSI- FELLOW/COFUND) and TRR80 of DFG. G. P. acknowledges support by the Humboldt Research Fellowship and by the Sonderforschungsbereich (SFB) 1143 project granted by DFG. F. H. thanks financial support through the grant SFB 1143.

\bibliographystyle{apsrev4-1}
\bibliography{Neutron_SrCuO2}

\end{document}